\documentclass{moriond}
\def\be{\begin{equation}}
\def\ee{\end{equation}}
\def\bea{\begin{eqnarray}}
\def\eea{\end{eqnarray}}

\newcommand{\Photo}{\includegraphics[height=35mm]{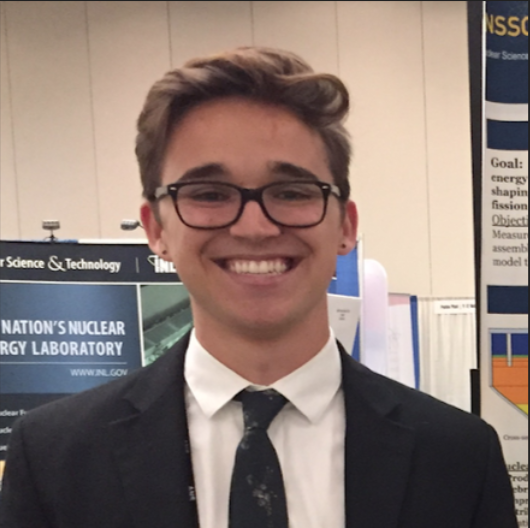}}

\begin{document}
\vspace*{4cm}
\title{Recent Results and Future Prospects from the STAR Beam Energy Scan Program}

\author{ Zachary Sweger on behalf of the STAR Collaboration}

\address{Department of Physics and Astronomy, University of California Davis, \\Davis, California 95616, USA}

\maketitle\abstracts{
The STAR experiment at Brookhaven National Laboratory has completed data taking for the second phase of the beam energy scan (BES-II) program, including in a fixed-target (FXT) mode. The BES-II program has collected high-statistics data on Au+Au collisions in the high baryon-density region of the QCD phase diagram. Together those data cover a wide range of per-nucleon center-of-mass energy from 3~GeV to 27~GeV. Recent results and anticipated analyses will be discussed along with implications for mapping the QCD phase diagram and its critical point.}

\section{Introduction}

The Solenoidal Tracker at RHIC (STAR) is a general-purpose detector providing precision particle identification and tracking. It is located at the Relativistic Heavy Ion Collider (RHIC) at Brookhaven National Laboratory (BNL). When running Au+Au collisions in collider mode, center-of-mass energies per nucleon pair of $7.7  \leq  \sqrt{s_{NN}} \leq 200$~GeV can be studied. 

\begin{figure*}
  \begin{center}
    \includegraphics[width=0.43\textwidth]{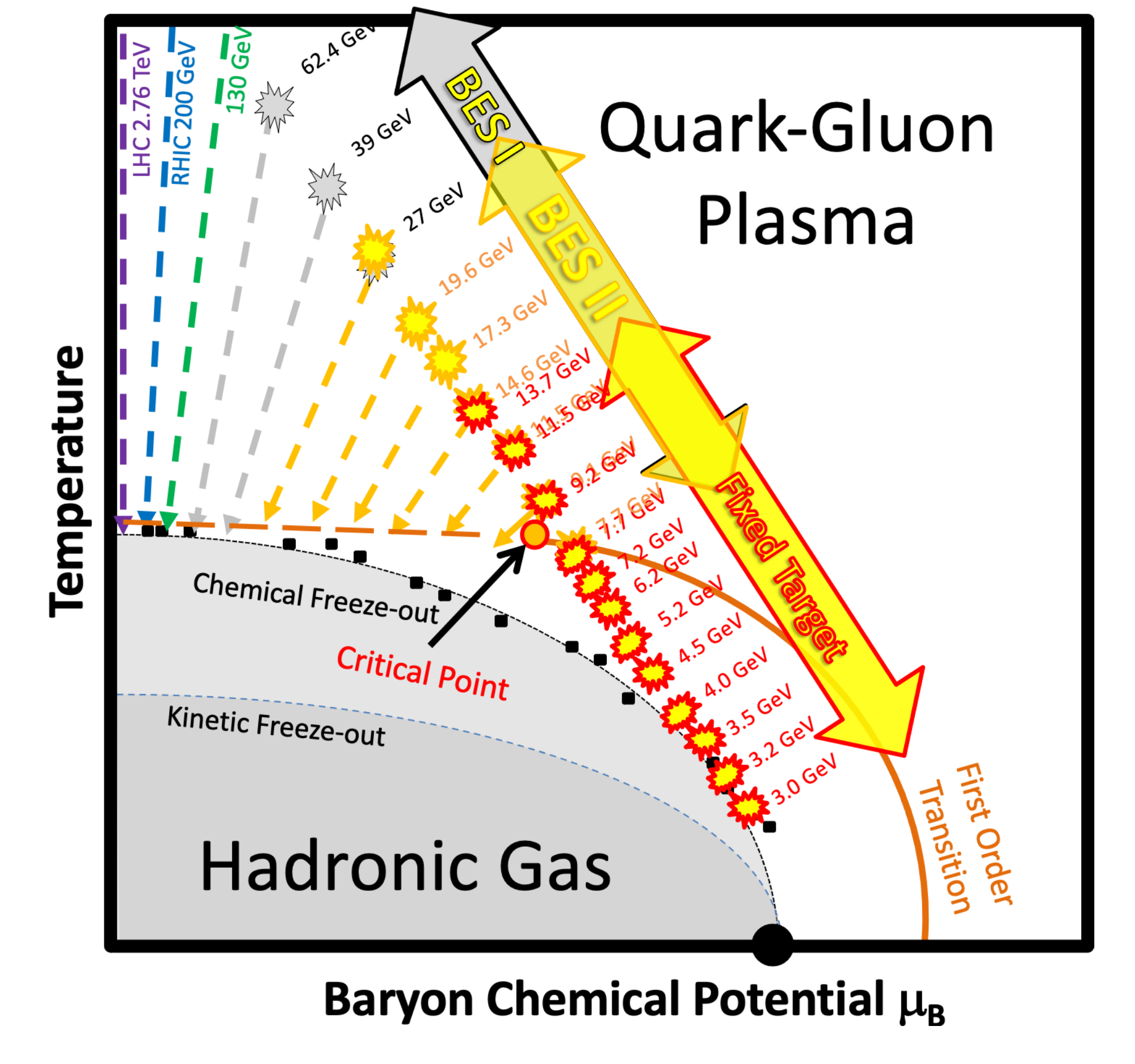}
    \caption{QCD phase diagram with BES-I, BES-II, and FXT coverages superimposed.~\protect\cite{Odyniec_2013}}
    \label{fig:qcd}
  \end{center}
\end{figure*}

In recent years, the STAR experiment has aimed to map the phases of QCD matter in Beam Energy Scan phase I and phase II, including the Fixed-Target Program (BES-I, BES-II, FXT). The phases of QCD matter are mapped as a function of temperature and baryon chemical potential ($\mu_B$) in Fig.~\ref{fig:qcd}. At low temperatures and $\mu_B$, partons are confined in a hadron gas. At high temperatures, quarks and gluons are deconfined and form a QGP. By colliding gold ions at varying energies, STAR is able to scan this phase space. The BES programs aim to map the size, shape, and temperature of the fireball produced in Au+Au collisions, as well as find the onset of deconfinement, signatures of a first-order phase transition, and a QCD critical point.

BES-I finished collecting data in 2011 and now BES-II and FXT have completed high-statistics data-taking at high-$\mu_B$. BES-II covers an energy range from $\sqrt{s_{NN}}=7.7$~GeV to $27$~GeV in the collider mode. To measure lower energies, the STAR utilized the FXT set up in 2018, which extends the $\sqrt{s_{NN}}$ coverage down to $3$~GeV. The STAR now has FXT data at 9 energies from $\sqrt{s_{NN}}=3.0$~GeV to $7.7$~GeV. One challenge with the FXT data is the shifting acceptance with respect to midrapidity, which limits some analyses at the highest FXT energies. 

Analyses of the BES-II data are ongoing. I will highlight here several analyses currently underway to characterize the evolution of the fireball and QGP as a function of collision energy. I will discuss the relevance of these analyses to our understanding of the phase diagram, and note what to look for in the coming months to years as results are made public.

\section{Femtoscopy and a First-Order Phase Transition}

Femtoscopy refers to the use of interferometry to measure the size and shape of the fireball produced in heavy-ion collisions. This field was inspired by Hanbury-Brown-Twiss (HBT) inteferometry~\cite{hbt}, in which interference of radio waves from a distant astrophysical object could be used to infer the size of the object. This principle can be used to map not only giant astrophysical objects, but also the femtometer-scale fireball produced in Au+Au collisions at RHIC. We refer to the three dimensions of the fireball (or HBT radii) as $R_{\rm long}$, $R_{\rm out}$, $R_{\rm side}$~\cite{PhysRevC.103.034908},  which characterize the longitudinal (beam direction), outward (along expansion transverse to the beam axis), and sideways (transverse to $R_{\rm long}$ and $R_{\rm out}$) dimensions of the fireball.

By measuring pion pairs, a two-particle correlation function, $C(\vec{q})$, may be constructed for identical pions as a function of the difference in their momenta ($\vec{q}$). The correlation function can be parameterized in terms of $R_{\rm out}$, $R_{\rm side}$, and $R_{\rm long}$, such that the fireball shape can be extracted. HBT radii were measured in BES-I~\cite{bes1hbt} and now BES-II analyses are being published~\cite{PhysRevC.103.034908}. 

HBT radii may provide an insight into the nature of the QCD phase transition. A peak in $R_{\rm out}/R_{\rm side}$ as a function of the collision energy can be predicted to be a signature of the first-order phase transition~\cite{firstorderfemtoscopy}. The latent heat associated with a first-order transition extends the lifetime of the fireball during freeze-out, during which a mixed phase is present. This extended lifetime during the fireball's outward expansion has the effect of extending $R_{\rm out}$, leading to a non-monotonicity in $R_{\rm out}/R_{\rm side}$ with respect to the collision energy. A peak structure was observed in BES-I near $\sqrt{s_{NN}}=20$~GeV, and the newest FXT measurement~\cite{PhysRevC.103.034908} at $\sqrt{s_{NN}}=4.5$~GeV is consistent with this picture. Further high-statistics measurements in BES-II and FXT will provide further insights on this first-order phase transition signature.

\section{Dileptons and the Fireball Temperature}

Leptons, which are transparent to the strong force, can be used to probe the temperature deep within the fireball up to the time at which chemical freezeout occurs. Dileptons are categorized based on their invariant mass region as either in the low-mass region (LMR, $M_{ll}<1.1$~GeV), intermediate-mass region (IMR, $1.1<M_{ll}<3$~GeV), or high-mass region (HMR, $M_{ll}>3$~GeV). LMR dileptons, resulting primarily from semileptonic meson decays, are able to measure the temperature of the fireball at chemical freeze-out when those mesons form. IMR dileptons, on the other hand, originate from thermal radiation of the fireball early in the collision. HMR dileptons come from hard processes and are not sensitive to temperature. Dileptons are therefore an ideal probe for measuring the temperature of the fireball both early in the collision and at chemical freeze-out. By studying these temperatures as a function of the collision energy, we may begin to understand where these temperatures diverge.

\begin{figure}
\begin{minipage}{0.47\linewidth}
\centerline{\includegraphics[width=1\linewidth]{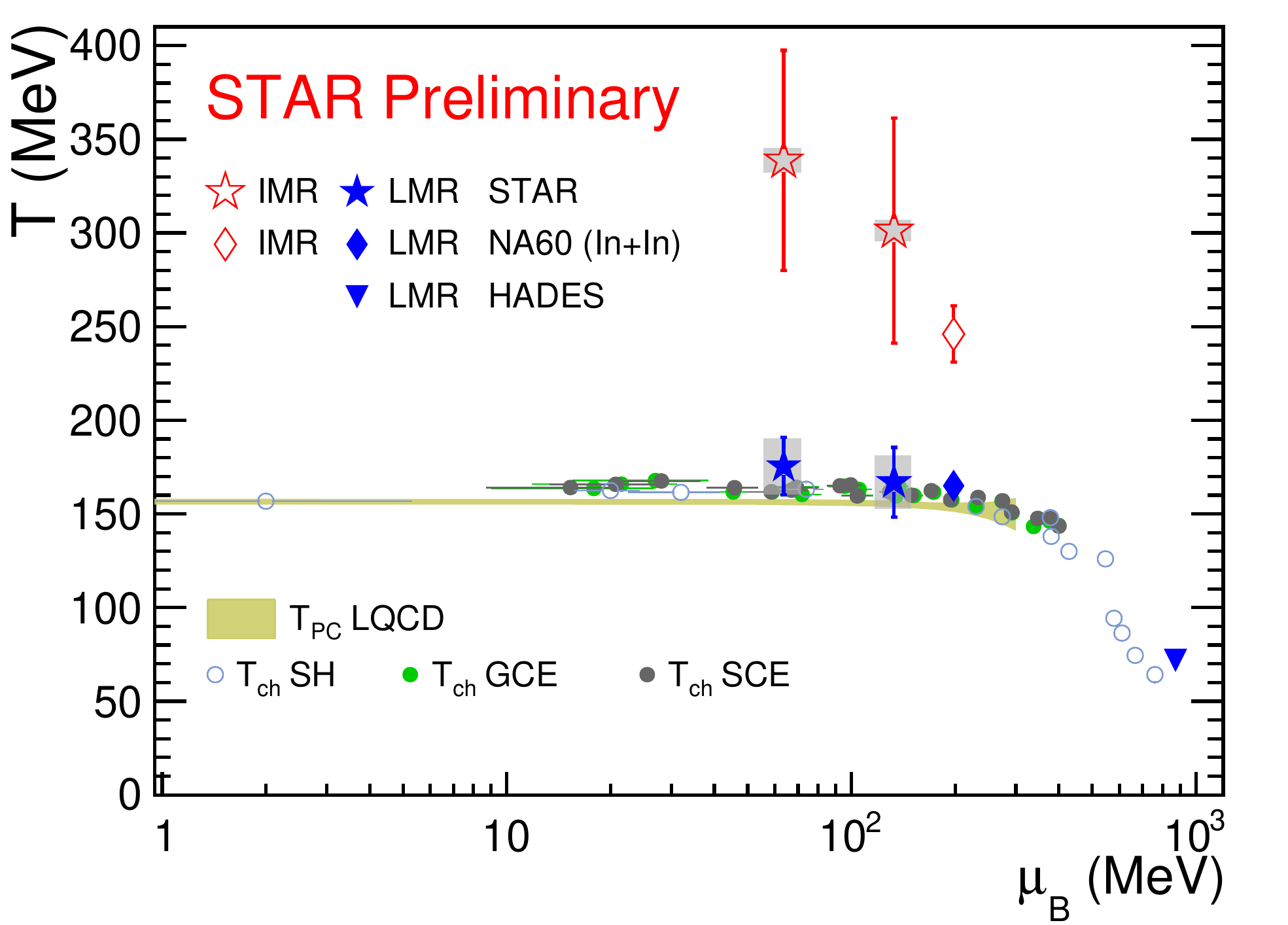}}
\end{minipage}
\hfill
\begin{minipage}{0.47\linewidth}
\centerline{\includegraphics[width=0.9\linewidth]{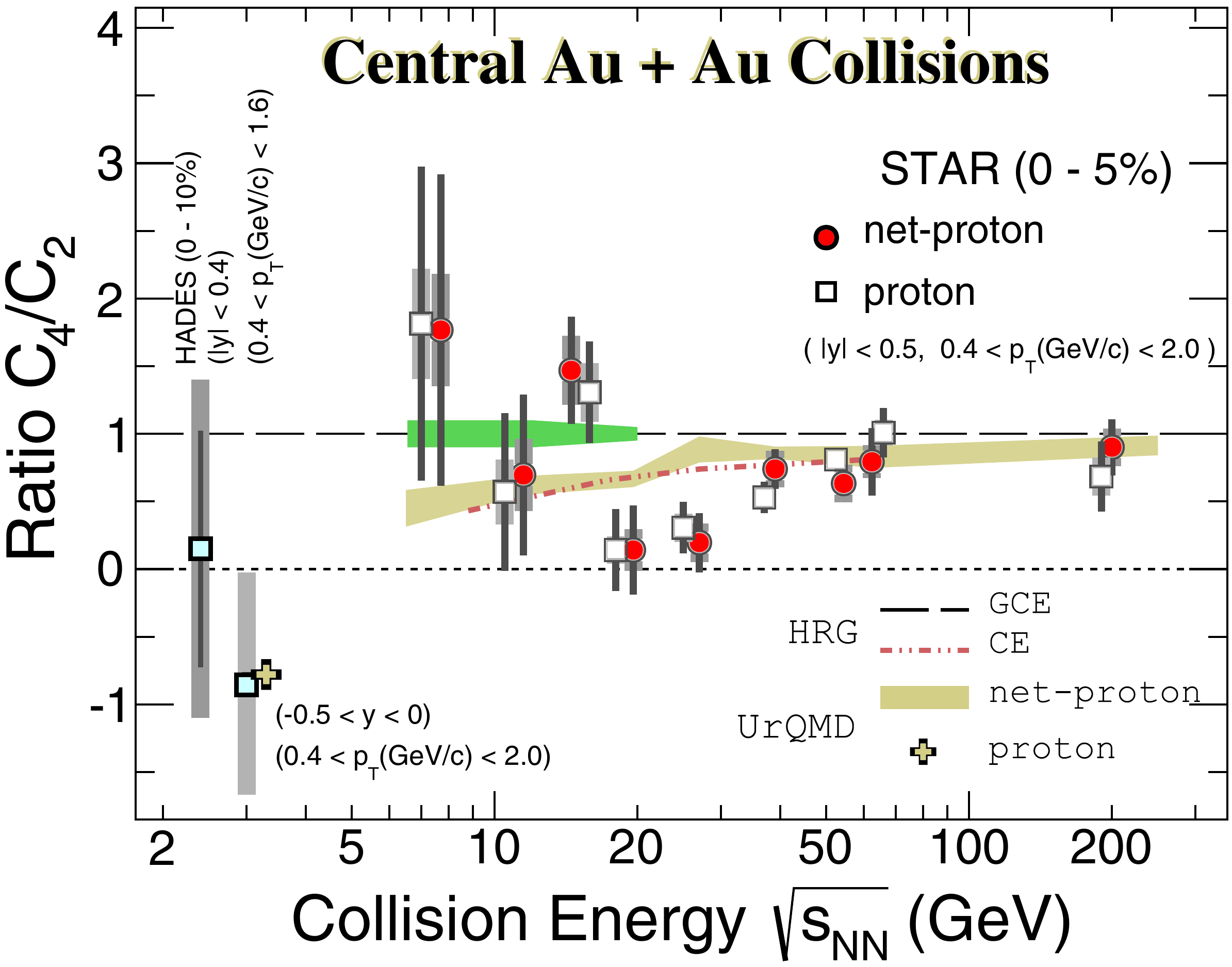}}
\end{minipage}
\hfill
\caption[]{IMR and LMR temperature measurements from STAR, NA60~\protect\cite{na60}, and HADES~\protect\cite{hades}, compared with chemical freeze-out obtained using a statistical-hadronization approach ($T_{ch}$ SH)~\protect\cite{Andronic2018}, a grand/strangeness canonical ensemble ($T_{ch}$ GCE/SCE)~\protect\cite{PhysRevC.96.044904}, and lattice QCD predictions ($T_{PC}$ LQCD)~\protect\cite{201915} (left). Proton kurtosis measurements~\protect\cite{heppelman} demonstrating non-monotonic deviations from the UrQMD baseline and return to baseline at 3~GeV (right).}
\label{fig:dielectrons}
\end{figure}

Recent results from STAR at $\sqrt{s_{NN}}=54.4$~GeV and $27$~GeV are pictured in Fig.~\ref{fig:dielectrons} (left) where the temperature extracted from the IMR diverges from that extracted from the LMR. This is comparable to the divergence in Fig.~\ref{fig:qcd} between the blast points representing the early fireball and the solid black points representing the state at chemical freeze-out. As the new data are processed, it will be interesting to see exactly where the initial-state and chemical freeze-out curves begin to diverge, and whether they follow the existing trends.

%\begin{figure*}
%  \begin{center}
%    \includegraphics[width=0.5\textwidth]{dielectrons.pdf}
%    \caption{IMR and LMR temperature measurements from STAR, NA60~\protect\cite{na60}, and HADES~\protect\cite{hades}, compared with chemical freeze-out obtained using a statistical-hadronization approach ($T_{ch}$ SH)~\protect\cite{Andronic2018} and using a grand/strangeness canonical ensemble ($T_{ch}$ GCE/SCE)~\protect\cite{PhysRevC.96.044904}. Lattice QCD predictions ($T_{PC}$ LQCD)~\protect\cite{201915} are shown as well.}
%    \label{fig:dielectrons}
%  \end{center}
%\end{figure*}

\section{Flow, Spectra, and the Onset of Deconfinement}

In non-central heavy-ion collisions, the overlap region of the two ions is an almond shape. The anisotropy of the collision geometry leads to pressure gradients within the fireball such that particles are also emitted anisotropically.  This is measured by taking a Fourier transform of the azimuthal distribution of detected particles, and extracting the coefficients $v_n$. The second coefficient $v_2$ is referred to as the elliptic flow and, for hadrons, scales proportionally with the number of constituent quarks (NCQ scaling). This tells us that early in the fireball while the flow dynamics are underway, the deconfined quarks themselves are flowing. For example, $v_2/3$ for protons matches $v_2/2$ for pions. 

The results from FXT $3.0$~GeV collisions have been published~\cite{2022137003}, in which, in stark contrast with previous measurements from $\sqrt{s_{NN}}=7.7$~GeV to $200$~GeV, NCQ scaling is broken. At $3.0$~GeV it is clear that pions and protons have very different elliptic flows, suggesting that the fireball produced in $\sqrt{s_{NN}}=3.0$~GeV collisions is made up of hadrons, not free partons.

Particle spectra, by measuring baryon stopping, also have a connection to deconfinement. Baryon stopping is a phenomenon in heavy-ion collisions in which baryons, initially confined in nuclei traveling at relativistic speeds, are detected at midrapidity. Thus a large number of protons have lost most of their momentum. It has been proposed that baryon stopping may be sensitive to the equation of state (EoS) of the fireball and may help resolve the order of the deconfinement phase transition~\cite{PhysRevC.87.064904}. Less stopping may be associated with a softening of the EoS, so we may search for a dip in the amount of stopping as a function of energy. Analysis of the FXT data has begun at $3.0$~GeV, and the results are consistent with previously-published results~\cite{benstalk}. The remaining data may illuminate the energy at which NCQ scaling breaks, and whether there is a decrease in baryon stopping, both of which may hint at the onset of deconfinement. 

\section{Cumulants of Proton-Number Distributions and the QCD Critical Point}
The last analysis I'll highlight here is the net-proton-number moments analysis, in which we count the number of protons detected in a given acceptance window subtracted by the number of antiprotons for each event and then calculate the $i$th-order cumulants of the distributions ($C_i$). Moments of distributions of conserved quantities are expected to undergo fluctuations if the fireball, while cooling, passes near the critical point. This analysis use fluctuations in proton number as a proxy for fluctuations in baryon number, and the trivial volume dependence of cumulants can be removed by taking ratios of different orders of cumulants. Model calculations predict a non-monotonic collision energy 
dependence of net-baryon $C_4/C_2$ in the vicinity of a critical point~\cite{stephanov}. A non-monotonicity was observed in BES-I with a $3.1\sigma$ significance~\cite{PhysRevLett.126.092301}. This exciting critical signature was part of the motivation for the BES-II and the FXT. The BES-II data covers the full range of the observed non-monotonicity, and the FXT data will be used to confirm the possible disappearance of the critical signature below 7.7~GeV.

The proton moments of the first FXT dataset at $\sqrt{s_{NN}}=3.0$~GeV have now been published~\cite{heppelman}. As shown in Fig.~\ref{fig:dielectrons} (right), the $C_4/C_2$ at this energy demonstrates a return to the non-critical (UrQMD) baseline, which could indicate that hadronic interactions are dominant at 3~GeV. The remaining task is to make high precision measurements from $\sqrt{s_{NN}}=3.2$~GeV to 27~GeV. This will allow us to see whether BES-I's non-monotonicity persists and determine if there is evidence of a critical point and its location.

\section{Conclusion}

STAR and RHIC are unique in their ability to map the QCD phase diagram with a large $\sqrt{s_{NN}}$ range and the ability to measure many observables. Analysis of recent data collected in STAR's BES-II and FXT are ongoing. With these high-statistics data sets, we hope to shed light on the temperature and size of the fireball produced in Au+Au collisions, as well as the onset of deconfinement and the location of the QCD critical point.

\end{document}